# RECENT RESULTS FROM K2K EXPERIMENT

C. Mariani (INFN Rome & University of Rome 'La Sapienza')

for K2K Collaboration

The latest results from K2K experiment are reported, with focus on muon neutrino oscillation. The results are based on data taken from June 1999 to February 2004, corresponding to $8.9 \times 10^{19}$ protons on target.

The KEK to Kamioka long-baseline neutrino experiment (K2K) is the first accelerator-based experiment with a distance of hundred kilometers. The intense nearly pure neutrino beam (98.2% $\nu_\mu$, 1.3% $\nu_e$, and 0.5% $\bar{\nu}_\mu$) has an average $L/E_\nu \approx 200 \text{km/GeV}^{-1}$ (250 km, $\langle E_\nu \rangle \approx 1.3$ GeV). The neutrino beam properties are measured just after the production, and the kinematics of parent pions are measured in situ to extrapolate the neutrino flux measured at the near detector to the expectation at the far detector. The neutrino beam energy spectrum and profile are measured by the near detector (ND) located 300m from the production target. The ND consists of two detector sets: a 1 kiloton water Cherenkov detector (1KT) and a fine grained detector (FGD) system. The far detector is SuperKamiokande (SK), a 50 kiloton water Cherenkov detector, located 250km from KEK. K2K focuses on the study of the existence of neutrino oscillation in $\nu_\mu$ disappearance that is observed in atmospheric neutrinos, and on the search for $\nu_\mu \rightarrow \nu_e$ oscillation with well understood flux and neutrino composition in the $\Delta m^2 \geq 2 \times 10^{-3}$ eV$^2$ region.

In this paper we will present the second result of K2K based on data taken from June 1999 to February 2004 corresponding to $8.9 \times 10^{19}$ protons on target (POT). From 1999 to 2001 (called K2K-I and SK-I), the inner detector surface of SK had 11,146 20-inch photo-multiplier tubes (PMts) covering 40% of the total area[3]. In K2K-I the FGD was comprised of a scintillating fiber and water target detector (SciFi)[5,6], a lead glass calorimeter, and a muon range detector (MRD)[5]. Starting from January 2003 (K2K-2 and SK-II), 19% of the SK inner detector is covered using 5182 PMts. In K2K-II the lead glass detector was removed from the FGD (K2K-IIa [$2.3\times10^{19}$POT]), and later, replaced by a fully active scintillator detector (SciBar)[6] (K2K-IIb [$1.9\times10^{19}$POT]). Adding K2K-II data almost double the statistics compared to the previous analysis[2].

The K2K neutrino beam[7,8], is a horn-focused wide band neutrino $\nu_\mu$ beam. The primary beam for K2K is 12 GeV kinetic energy protons from the KEK proton synchrotron (KEK-PS)[9,10]. Downstream the horn system, before the 200m long decay volume where the pions decay to $\nu_\mu$ and muons, a gas-Cherenkov detector (PIMON)[4] is occasionally put in the beam to measure the kinematics of the pions after their production and subsequent focusing. After the decay volume, an iron and concrete beam dump stops essentially all charged particles except muons with energy greater than 5.5 GeV. Downstream the dump there is a muon monitor (MUMON)[4] consisting of a segmented ionization chamber and an array of silicon pad detectors. This monitors the residual muons spill-by-spill to check beam centering and muon yields. The ionization chambers consist of 5 cm wide strips covering $2 \times 2$ m$^2$ area with separate planes for horizontal and vertical read-out. The 26 silicon pads, each of about 10 cm$^2$ are distributed through a 3x3 m$^2$ area. The neutrino beam direction is finally monitored using neutrino events in the MRD. It is stable, within 1 mrad throughout the entire experimental period. The same events are used to confirm that the energy spectrum is stable.

The 1KT detector is used to measure the expected total number of interactions at SK. The 1KT detector uses the same water target as SK and the uncertainties in the neutrino cross section cancel. The event selection criteria and the 25 ton fiducial volume is the same as in previous analysis[2]. In order to estimate the $E_\nu$ spectrum along with the other near detectors, we use the subset of fully contained events with only one Cherenkov ring, which is judged to be a muon based on the distribution of Cherenkov light (1 ring μ-like events). For these events we measure both the momentum ($p_\mu$) and angle ($\theta_\mu$), from which we estimate the neutrino energy spectrum (the largest uncertainty of +2/-3% comes from the overall energy scale). Neutrino interaction is also studied with the SciFi and SciBar detectors.

The SciFi detector is made of layers of scintillating fibers between tanks of aluminum filled with water. The fiducial volume is 5.6 tons. The K2K-I analysis includes CC events with the muon reaching the MRD detector and also events in which the muon track stops in the lead glass, with momentum as low as 400MeV/c, significantly lowering the energy threshold compared to previous analysis[2].

The SciBar detector[6] consists of 14848 extruded scintillators strips read out by wavelength shifting fibers and multi-anode PMts. Both timing and charge of the PMT outputs are recorded. Strips with dimensions of 1.3 x 2.5 x 300 cm$^3$ are arranged in 64 layers. Each layer consists of two planes to measure horizontal and vertical position. The 4 radiation lengths of the scintillator are followed downstream by a layer of 11 radiation lengths consisting of two projective planes of lead and fiber "spaghetti" calorimeter which provide event containment and additional energy measurement. The scintillator also acts as the neutrino interaction target; SciBar is a fully active detector and has high efficiency for low momentum particles. Since the target material is different from water, differences due to nuclear effect and the related systematic uncertainty are taken into account in the measurements of cross-section.

In SciBar, tracks which traverse at least three layers (about 8 cm) are reconstructed. The reconstruction efficiency for an isolated track longer than 10 cm is 99%. We select charged current (CC) events by requiring at least one of the tracks start from the 9.38 ton fiducial volume and extend to the MRD. With this requirement, the $p_\mu$ threshold is 450MeV/c. The $p_\mu$ scale uncertainty, the $p_\mu$ resolution and the $\theta_\mu$ resolution are 2.7%, 80MeV/c and 1.6°, respectively. The efficiency for a second, short track is lower than that for a muon track mainly due to the overlap with the primary track. This efficiency smoothly goes up from the threshold (8 cm, corresponding to 450 MeV/c proton) and reaches 90% at 30cm (670MeV/c for proton). For SciFi and SciBar, we select events in which one or two tracks are reconstructed. For two tracks events we use kinematics information to discriminate between quasi elastic interactions (QE) and non quasi elastic interactions (non-QE). The direction of the recoil proton can be predicted from the $p_\mu$ and $q_\mu$ assuming a QE interaction. If the difference between the predicted and observed direction of the second track is lower than 25° the interaction is assumed to be QE. Interactions for which this difference is large (more than 30° for SciFi and more than 25° for SciBar) are put into the non-QE sample. In total we select seven exclusive samples: 1 ring μ-events in 1 KT, 1 track in SciFi and SciBar, 2 tracks QE and 2 tracks non-QE in SciBar and SciFi.

We measure the $E_\nu$ spectrum at the ND from this samples by fitting simultaneously all two-dimensional distributions of $p_\mu$ and $\theta_\mu$ with a baseline Monte Carlo expectation[2]. We obtain the cross section ratio of non-QE to QE interaction ($R_{nqe}$) relative to our MC simulation. We observe a deficit of forward going muon in all ND data samples compared to the MC. To avoid bias due to this, we perform the $E_\nu$ spectrum fit using only data with $\theta_\mu$ >20 (10) degrees for 1KT (SciFi and SciBar). The $\chi^2$ value for the best fit is 538.5 for 479 degrees of freedom (DOF). The resulting $E_\nu$ spectrum and its error are summarized in Tab. I, while the best fit for $R_{nqe}$ is 0.95. Possible sources of the forward deficit could be the amount of resonant pion production and coherent pion production at low $q^2$ in our MC. We modify the MC simulation used to take this into account. For resonant pion we suppress the cross section by $q^2/A$ for $q^2 < A$ and leave it unchanged for $q^2 > A$. From a fit to the SciBar 2-track non-QE sample, A is $0.10 \pm 0.03$ $(GeV/c)^2$. Alternatively, if we assume that this deficit is due to coherent pion production, we find the observed distribution is reproduced best with no coherent pion.

| $E_\nu$ (GeV) | $\phi_{ND}$ | $\Delta(\phi_{ND})$ [%] | $\Delta(F/N)$ [%] | $\Delta(\varepsilon_{SK-I})$ [%] | $\Delta(\varepsilon_{SK-II})$ [%] |
|---|---|---|---|---|---|
| 0.0 – 0.5 | 0.032 | 46 | 2.6 | 3.7 | 4.5 |
| 0.5 – 0.75 | 0.32 | 8.5 | 4.3 | 3.0 | 3.2 |
| 0.75 – 1.0 | 0.73 | 5.8 | 4.3 | 3.0 | 3.2 |
| 1.0 – 1.5 | $\equiv 1$ | - | 4.9 | 3.3 | 8.2 |
| 1.5 – 2.0 | 0.69 | 4.9 | 10 | 4.9 | 7.8 |
| 2.0 – 2.5 | 0.34 | 6.0 | 11 | 4.9 | 7.4 |
| 2.5 – 3.0 | 0.12 | 13 | 12 | 4.9 | 7.4 |
| 3.0 – | 0.049 | 17 | 12 | 4.9 | 7.4 |

TABLE I: The $E_\nu$ spectrum fit results. $\phi_{ND}$ is the best fit value of relative flux for each $E_\nu$ bin to the 1.0–1.5 GeV bin. The uncertainties in $\phi_{ND}$, F/N ratio, and reconstruction efficiencies for SK-I and SK-II are also shown.

Considering both possibilities mentioned above, we fit the parameter $R_{nqe}$ again and check the agreement with the data. The $E_\nu$ spectrum is kept fixed at the values already obtained in the first step, but now we use data at all angles. The best fit value for $R_{nqe}$ is 1.02 (1.06) with $\chi^2$/DOF of 638.1/609 (667.1/606) when we suppress resonant pion (eliminate the coherent pion). The $p_\mu$ and $\theta_\mu$ distribution from all detectors are well reproduced for both cases with reasonable $\chi^2$ as show in fig.1. If we repeat the fit with the $E_\nu$ spectrum free, the results are still consistent with the first step. These results do not allow to surely identify the source of the observed deficit in low $q^2$ region. An additional systematic error of 0.1 is assigned to $R_{nqe}$ to take into account the dependency from the low $q^2$ model. For the oscillation analysis presented in this paper, we choose to suppress the resonance pion production in MC simulation and when we determine the central value of $R_{nqe}$. However, we find that the final oscillation results do not change if we instead choose to eliminate coherent pion production or if we use our MC without any correction.

For the oscillation analysis, events in SK are selected based on timing information from the global positioning system. From out of time events the background level from atmospheric neutrinos is estimated to be 2x10$^{-3}$ events. For K2K I+II there are 107 events in the 22.5kTon fiducial volume that are fully contained, have no energy seen in the outer detector, and have at least 30MeV deposited energy in the inner detector.

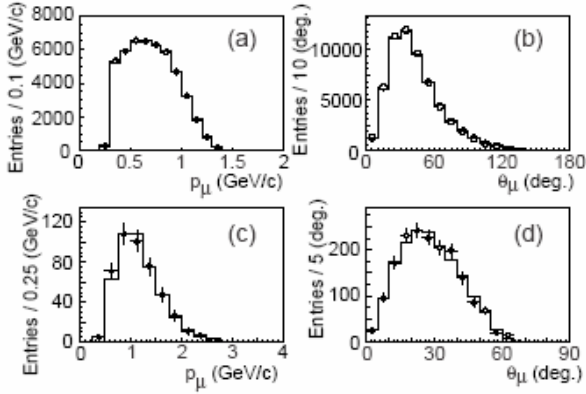

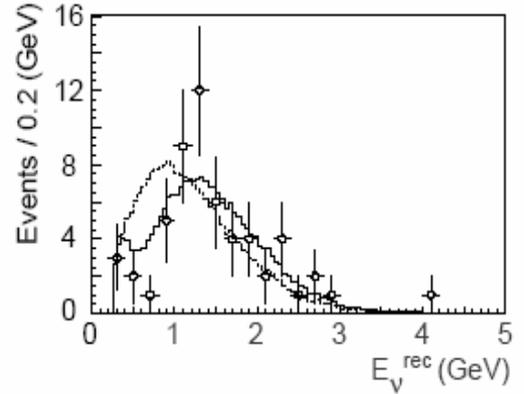

Figure 1: A selection of muon momentum (p$_\mu$) and direction ($\theta_\mu$) distributions: (a) the p$_\mu$ distribution of 1KT fully contained 1-ring μ-like sample, (b) 1KT $\theta_\mu$ for the same sample, (c) SciFi p$_\mu$ for 2-track QE sample, and (d) SciBar $\theta_\mu$ for 2-track non-QE sample. Open circles represent data, while histograms are MC predictions using the best fit E$_\nu$ spectrum and suppression of the resonant pion production.

Figure 2: The reconstructed E$_\nu$ distribution for the SK 1-ring μ-like sample. Points with error bars are data. The solid line is the best fit spectrum. The dashed line is the expected spectrum without oscillation. These histograms are normalized to the number of events observed (57).

The expected number of SK without oscillation is $151^{+12}_{-10}$ (syst). The major contributions to the errors come from the uncertainties in the F/N ratio (5.1%); the latter is dominated by the uncertainty in the fiducial volume due to vertex reconstruction at both SK and 1KT. We reconstruct the neutrino energy assuming CC-QE kinematics, from p$_\mu$ and $\theta_\mu$ for the 57 events in the 1-ring μ-like sample subset of SK data. With these we measure the energy spectrum distortion caused by neutrino oscillation. The detector systematics of SK-I and SK-II are slightly different because of the change in the PMts number in the inner detector. The main contribution to the systematic error of the oscillation analysis based on the energy spectrum is the energy scale uncertainties: 2.0% for SK-I and 2.1% for SK-II.

Uncertainties for the ring counting and particles identification are estimated using the atmospheric neutrino data sample and MC simulation. A two flavor neutrino oscillation analysis, with $\nu_\mu$ disappearance, is performed using a maximum-likelihood method. The oscillation parameters, ($\sin^2 2\theta$, $\Delta m^2$), are estimated by maximizing the product of the likelihood for the observed number of FC events ($L_{num}$) and that for the shape of the $E_\nu^{rec}$ spectrum ($L_{shape}$). $L_{num}$ is a binned likelihood for the two dimensional $p_\mu$ - $\theta_\mu$ distribution, constructed from poissonian probability density function for each bin. The PDF for $L_{shape}$ is the expected $E_\nu^{rec}$ distribution at SK, which is estimated from MC simulation. The PDFs are defined for K2K-I and K2K-II separately. The systematic uncertainties due to the following sources are taken into account in the PDFs: the $E_\nu$ measured spectrum by ND, the Far/Near ratio, the reconstruction efficiency and absolute energy scale of SK, the ratio of neutral current to CC-QE cross section, the ratio of CC non-QE to CC-QE cross section and the overall normalization. The expected distributions depend on the systematic uncertainty parameters, which are assumed to follow a Gaussian distribution[1]. A constraint term ($L_{syst}$) is multiplied with the likelihood for each of these systematics, and $L_{num}$x$L_{shape}$x$L_{syst}$ is maximized during the fit. The total number of fit parameters is 34. The best fit point within the physical region is ($\sin^2 2\theta=1.0$, $\Delta m^2=2.8 \times 10^{-3}$ eV$^2$); the expected number of events at this point is 103.8, which agrees well with the 107 observed events. The best fit $E_\nu$ distribution is shown in fig.2. The consistency between the observed and fit $E_\nu$ distribution is checked using Kolmogorov-Smirnov (KS) test. For the best fit parameters. The KS probability is 36% while that for the non oscillation hypothesis is 0.08%. A highest likelihood point ($\sin^2 2\theta=1.5$, $\Delta m^2=2.2 \times 10^{-3}$ eV$^2$) is found outside the physical region. The probability that we would get $\sin^2 2\theta > 1.5$ if the true parameters are our best fit, is 13%, based on MC virtual experiments. For the rest of the paper we will only refer as best fit to the best fit in physical region. The possibility that the observation is due to a statistical fluctuation instead of neutrino oscillation is estimated by computing the likelihood ratio between the non oscillation case and the best fit point. Without oscillation the probability of our result is 0.0050% (4.0σ). When only normalization (shape) information is used, the probability is 0.26% (0.74%). Allowed regions for the oscillation parameters are evaluated by calculating the likelihood ratio of each point to the best fit point and are drawn in fig. 3. The 90% C.L. contour crosses the $\sin^2 2\theta = 1$ axis at $\Delta m^2 = 1.9$ and $3.6 \times 10^{-3}$ eV$^2$. The oscillation parameters from the $E_\nu$ spectrum distortion alone, or the total event analysis also agree.

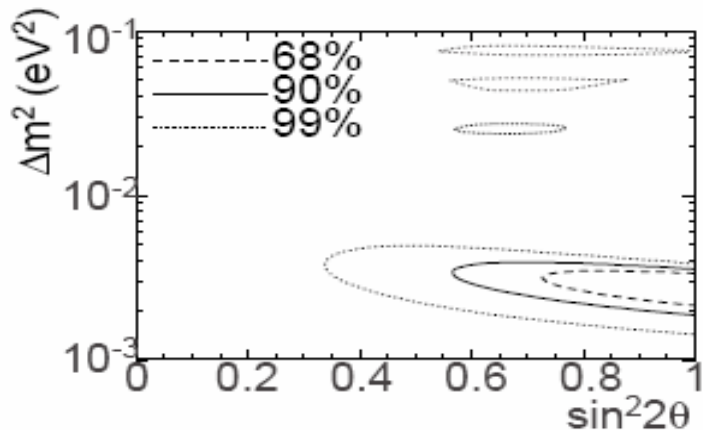

Figure 3: Allowed regions of oscillation parameters. Dashed, solid and dot-dashed lines are 68.4%, 90% and 99% C.L. contours, respectively.

We also performed the analysis eliminating the coherent pion production instead of suppressing the resonant pions and the allowed regions are unchanged; the end result is the same. In conclusion using accelerator produced neutrinos, we confirm at 4σ that oscillation of muon neutrino occurs with the same parameters observed in atmospheric neutrino measurements[1].